\documentclass[aps,prc,preprint,tightenlines,groupedaddress,nofootinbib,
noshowpacs,preprintnumbers,amsmath,amssymb,superscriptaddress]{revtex4}
\usepackage{graphicx}
\usepackage{dcolumn}
\usepackage{bm}
\usepackage{amsmath}
\begin{document}

\title{Neutron Rich Nuclei in Heaven and Earth}
\author{J. Piekarewicz}
\affiliation{Department of Physics, Florida State 
             University, Tallahassee, FL 32306, USA}

\begin{abstract}
The nucleus of ${}^{208}$Pb --- a system that is 18 order of
magnitudes smaller and 55 orders of magnitude lighter than a neutron
star --- may be used as a miniature surrogate to establish important
correlations between its neutron skin and several neutron-star
properties. Indeed, a nearly model-independent correlation develops
between the neutron skin of $^{208}$Pb and the liquid-to-solid
transition density in a neutron star.  Further, we illustrate how a
measurement of the neutron skin in ${}^{208}$Pb may be used to place
important constraints on the cooling mechanism operating in neutron
stars and may help elucidate the existence of quarks stars.
\end{abstract}
\maketitle 

\section{Introduction}
\label{Sec:Introduction}

It is an extrapolation of 18 orders of magnitude from the neutron
radius of a heavy nucleus --- such as $^{208}$Pb with a neutron radius
of $R_{n}\!\approx\!5.7$~fm --- to the approximately 10~km radius of a
neutron star. Yet both radii depend on our incomplete knowledge of the
equation of state of neutron-rich matter. That strong correlations
arise among objects of such disparate sizes is not difficult to
understand. Heavy nuclei develop a neutron-rich skin as a result of
its large neutron excess ({\it e.g.,} $N/Z\!=\!1.54$ in $^{208}$Pb)
and because the large Coulomb barrier reduces the proton density at
the surface of the nucleus. Thus the thickness of the neutron skin
depends on the pressure that pushes neutrons out against surface
tension. As a result, the greater the pressure, the thicker the
neutron skin~\cite{Brown:2000}. Yet it is this same pressure that
supports a neutron star against gravitational
collapse~\cite{Lattimer:2000nx,Steiner:2004fi}. Thus models with
thicker neutron skins often produce neutron stars with larger
radii~\cite{Horowitz:2001ya}.

The above discussion suggests that an accurate and model-independent
measurement of the neutron skin of even a single heavy nucleus
may have important implications for neutron-star properties.
Attempts at mapping the neutron distribution have traditionally relied
on strongly-interacting probes. While highly mature and successful, it
is unlikely that the hadronic program will ever attain the precision
status that the electroweak program enjoys. This is due to the large
and controversial uncertainties in the reaction
mechanism~\cite{Ray:1985yg,Ray:1992fj}. The mismatch in our knowledge
of the proton radius in ${}^{208}$Pb relative to that of the neutron
radius provides a striking example of the current situation: while the
charge radius of ${}^{208}$Pb is known to better than
0.001~fm~\cite{Fricke:1995}, realistic estimates place the uncertainty
in the neutron radius at about 0.2 fm~\cite{Horowitz:1999fk}.

The enormously successful parity-violating program at the Jefferson
Laboratory~\cite{Aniol:2005zf,Aniol:2005zg} provides an attractive
electroweak alternative to the hadronic program. Indeed, the Parity
Radius Experiment (PREX) at the Jefferson Laboratory aims to measure
the neutron radius of $^{208}$Pb accurately (to within $0.05$~fm) and
model independently via parity-violating electron
scattering~\cite{Horowitz:1999fk}. Parity violation at low momentum
transfers is particularly sensitive to the neutron density because the
$Z^0$ boson couples primarily to neutrons. Moreover, the
parity-violating asymmetry, while small, can be interpreted with as
much confidence as conventional electromagnetic scattering
experiments. PREX will provide a unique observational constraint on
the thickness of the neutron skin of a heavy nucleus. We note that
since first proposed in 1999, many of the technical difficulties
intrinsic to such a challenging experiment have been met. For example,
during the recent activity at the Hall A Proton Parity Experiment
(HAPPEX), significant progress was made in controlling helicity
correlated errors~\cite{Michaels:2005}. Other technical problems are
currently being solved --- such as the designed of a new septum magnet
--- and a specific timeline has been provided to solve all remaining
problems within the next two years~\cite{Michaels:2005}.

Our aim in this contribution is to report on some of our recent
results that examine the correlation between the neutron skin of
${}^{208}$Pb and various neutron-star
properties~\cite{Horowitz:2000xj,Horowitz:2001ya, Horowitz:2002mb}.
In particular, we examine the consequences of a ``softer'' equation 
of state that is based on a new accurately calibrated relativistic
parameter set that has been constrained by both the ground state
properties of finite nuclei and their linear response. Further,
results obtained with this new parameter set --- dubbed
``FSUGold''~\cite{Todd-Rutel:2005fa} --- will be compared against 
the NL3 parameter set of Lalazissis, Konig, and
Ring~\cite{Lalazissis:1996rd,Lalazissis:1999} that, while highly
successful, predicts a significantly stiffer equation of state.

\section{Formalism}
\label{Sec:Formalism}

The starting point for the calculation of the properties of finite
nuclei and neutron stars is an effective field-theory model based
on the following Lagrangian density:
\begin{widetext}
\begin{eqnarray}
&&
{\cal L}_{\rm int} =
\bar\psi \left[g_{\rm s}\phi   \!-\!
         \left(g_{\rm v}V_\mu  \!+\!
    \frac{g_{\rho}}{2}{\mbox{\boldmath $\tau$}}\cdot{\bf b}_{\mu}
                               \!+\!
    \frac{e}{2}(1\!+\!\tau_{3})A_{\mu}\right)\gamma^{\mu}
         \right]\psi \nonumber \\
                   && -
    \frac{\kappa}{3!} (g_{\rm s}\phi)^3 \!-\!
    \frac{\lambda}{4!}(g_{\rm s}\phi)^4 \!+\!
    \frac{\zeta}{4!}
    \Big(g_{\rm v}^2 V_{\mu}V^\mu\Big)^2 \!+\!
    \Lambda_{\rm v}
    \Big(g_{\rho}^{2}\,{\bf b}_{\mu}\cdot{\bf b}^{\mu}\Big)
    \Big(g_{\rm v}^2V_{\mu}V^\mu\Big) \;.
\label{Lagrangian}
\end{eqnarray}
\end{widetext}
The Lagrangian density includes an isodoublet nucleon field ($\psi$)
interacting via the exchange of two isoscalar mesons --- a scalar
($\phi$) and a vector ($V^{\mu}$) --- one isovector meson ($b^{\mu}$),
and the photon ($A^{\mu}$)~\cite{Serot:1984ey,Serot:1997xg}. In
addition to meson-nucleon interactions, the Lagrangian density is
supplemented by four nonlinear meson interactions, with coupling
constants denoted by $\kappa$, $\lambda$, $\zeta$, and 
$\Lambda_{\rm v}$. The first three of these terms are responsible 
for a softening of the equation of state of symmetric nuclear matter
at both normal and high densities~\cite{Mueller:1996pm}. In particular,
the cubic ($\kappa$) and quartic ($\lambda$) scalar self-energy terms 
are needed to reduce the compression modulus of symmetric nuclear 
matter, in accordance to measurements of the giant monopole resonance 
in medium to heavy nuclei~\cite{You99_PRL82}. In turn, $\omega$-meson 
self-interactions ($\zeta$) are instrumental for the softening of the 
equation of state at high density --- thereby affecting primarily the 
limiting masses of neutron stars~\cite{Mueller:1996pm}. Finally, the 
last of the coupling constants ($\Lambda_{\rm v}$) induces 
isoscalar-isovector mixing and has been added to tune the poorly-known 
density dependence of the symmetry 
energy~\cite{Horowitz:2000xj,Horowitz:2001ya}. As a result of the
strong correlation between the neutron radius of heavy nuclei and
the pressure of neutron-rich matter~\cite{Brown:2000,Furnstahl:2001un}, 
the neutron skin of a heavy nucleus is highly sensitive to changes 
in $\Lambda_{\rm v}$.

\section{Results}
\label{Sec:Results}

\begin{table*}
\begin{tabular}{|l|cccccccc|}
 \hline
 Model & $m_{\rm s}$  & $g_{\rm s}^2$ & $g_{\rm v}^2$ & $g_{\rho}^2$
       & $\kappa$ & $\lambda$ & $\zeta$ & $\Lambda_{\rm v}$\\
 \hline
 \hline
 NL3     & 508.1940 & 104.3871  & 165.5854 &  79.6000
         & 3.8599 & $-$0.0159 & 0.0000 & 0.0000   \\
 \hline
 FSUGold & 491.5000 & 112.1996  & 204.5469 & 138.4701
         & 1.4203 & $+$0.0238 & 0.0600 & 0.0300   \\
\hline
\end{tabular}
\caption{Model parameters used in the calculations. The parameter
$\kappa$ and the inverse scalar range $m_{\rm s}$ are given in MeV.
The nucleon, omega, and rho masses are kept fixed at $M\!=\!939$~MeV,
$m_{\omega}\!=\!782.5$~MeV, and $m_{\rho}\!=\!763$~MeV, respectively.}
\label{Table0}
\end{table*}

\begin{figure}[ht]
\vspace{0.50in}
\includegraphics[width=5in,angle=0]{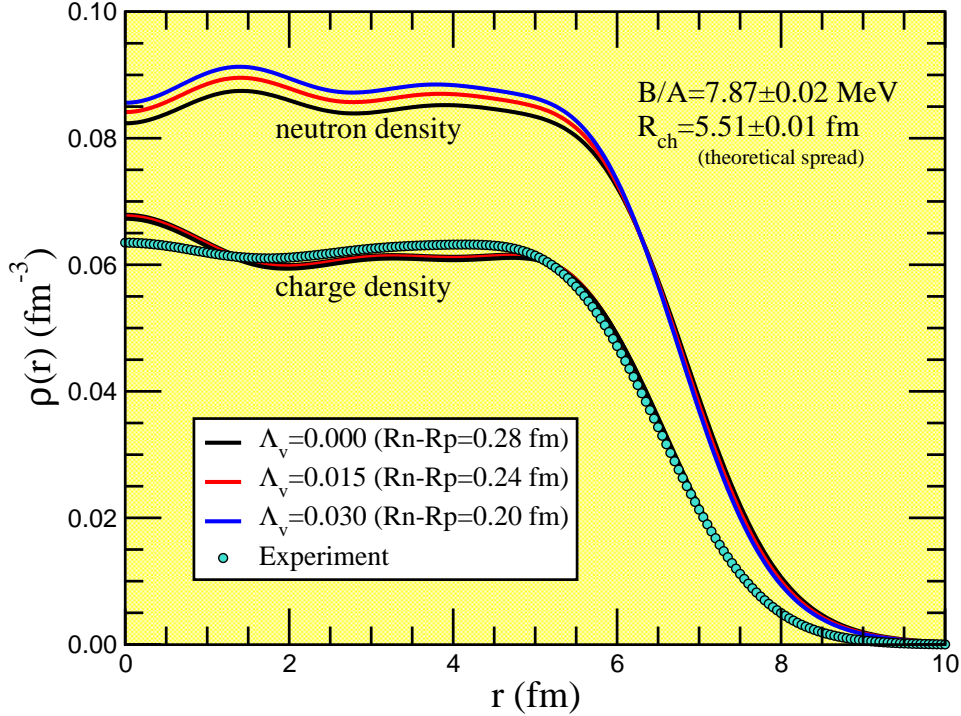}
\caption{Proton (charge) and neutron (point) densities for ${}^{208}$Pb
         using a variety of values for the isoscalar-isovector coupling
         constant $\Lambda_{\rm v}$.}
\label{Fig1}
\end{figure}

In Table~\ref{Table0} we list the various mass-parameters and coupling
constants for both of the models (NL3 and FSUGold) employed in the
text.  Further, in Table~\ref{Table1} a comparison is made between the
very successful NL3
parametrization\cite{Lalazissis:1996rd,Lalazissis:1999}, FSUGold, and
(when available) experimental data. Finally, in Fig.~\ref{Fig1} we
display the charge and neutron density of ${}^{208}$Pb as a function
of the isoscalar-isovector parameter $\Lambda_{\rm v}$. Note that
while $\Lambda_{\rm v}$ modifies the neutron radius of a heavy
nucleus, it does so without affecting those ground-state properties
that are well constrained by experiment, such as binding energies and
charge radii.

  \begin{table}
  \begin{tabular}{|c|c|c|c|c|}
    \hline
    Nucleus & Observable & Experiment & NL3 & FSUGold \\
    \hline
    \hline
    ${}^{40}$Ca & $B/A$~(MeV)            & $8.55$ & $\phantom{-}8.54$ & $\phantom{-}8.54$  \\
                & $R_{\rm ch}$~(fm)      & $3.45$ & $\phantom{-}3.46$ & $\phantom{-}3.42$  \\
                & $R_{n}\!-\!R_{p}$~(fm) &   ---  & $-0.05$           & $-0.05$            \\
    \hline
    ${}^{48}$Ca & $B/A$~(MeV)            & $8.67$ & $\phantom{-}8.64$ & $\phantom{-}8.58$  \\
                & $R_{\rm ch}$~(fm)      & $3.45$ & $\phantom{-}3.46$ & $\phantom{-}3.45$  \\
                & $R_{n}\!-\!R_{p}$~(fm) &   ---  & $\phantom{-}0.23$ & $\phantom{-}0.20$  \\
    \hline
    ${}^{90}$Zr & $B/A$~(MeV)            & $8.71$ & $\phantom{-}8.69$ & $\phantom{-}8.68$  \\
                & $R_{\rm ch}$~(fm)      & $4.26$ & $\phantom{-}4.26$ & $\phantom{-}4.25$  \\
                & $R_{n}\!-\!R_{p}$~(fm) &   ---  & $\phantom{-}0.11$ & $\phantom{-}0.09$  \\
    \hline
   ${}^{116}$Sn & $B/A$~(MeV)            & $8.52$ & $\phantom{-}8.48$ & $\phantom{-}8.50$  \\
                & $R_{\rm ch}$~(fm)      & $4.63$ & $\phantom{-}4.60$ & $\phantom{-}4.60$  \\
                & $R_{n}\!-\!R_{p}$~(fm) &   ---  & $\phantom{-}0.17$ & $\phantom{-}0.13$  \\
    \hline
   ${}^{132}$Sn & $B/A$~(MeV)            & $8.36$ & $\phantom{-}8.37$ & $\phantom{-}8.34$  \\
                & $R_{\rm ch}$~(fm)      &   ---  & $\phantom{-}4.70$ & $\phantom{-}4.71$  \\
                & $R_{n}\!-\!R_{p}$~(fm) &   ---  & $\phantom{-}0.35$ & $\phantom{-}0.27$  \\
    \hline
   ${}^{208}$Pb & $B/A$~(MeV)            & $7.87$ & $\phantom{-}7.88$ & $\phantom{-}7.89$  \\
                & $R_{\rm ch}$~(fm)      & $5.50$ & $\phantom{-}5.51$ & $\phantom{-}5.52$  \\
                & $R_{n}\!-\!R_{p}$~(fm) &   ---  & $\phantom{-}0.28$ & $\phantom{-}0.21$  \\
    \hline
  \end{tabular}
 \caption{Experimental data for the binding energy per nucleon and the
          charge radii for the magic nuclei used in the least square
          fitting procedure. In addition, predictions are displayed
          for the neutron skin of these nuclei.}
  \label{Table1}
 \end{table}

While the agreement (at the 1\% level or better) between NL3 and
experiment is satisfactory --- and this agreement extends all over the
periodic table~\cite{Lalazissis:1999} --- incorporating additional
terms into the density functional is demanded, not by ground state
data, but rather by their linear response. Indeed, as argued in
Refs.~\cite{Piekarewicz:2002jd,Piekarewicz:2003br} the success of the
NL3 set in reproducing the breathing mode in ${}^{208}$Pb is
accidental, as it results from a combination of both a stiff equation
of state for symmetric nuclear matter and a stiff symmetry
energy. That this is indeed the case may be appreciated by examining
the breathing mode in ${}^{90}$Zr --- a nucleus with well-developed
monopole strength but rather insensitive to the symmetry energy ---
and the energy of the isovector giant dipole resonance (IVGDR) in
${}^{208}$Pb --- an observable sensitive to the density dependence of
the symmetry energy.  In Table~\ref{Table2} relativistic random phase
approximation (RPA) results for the centroid energies of the giant
monopole resonance (GMR) in ${}^{208}$Pb and ${}^{90}$Zr, as well as
the IVGDR (peak energy) in ${}^{208}$Pb are reported.  These
small-amplitude modes represent the linear response of the mean field
ground state to a variety of probes~\cite{Rit93_PRL70,You99_PRL82}. As
alluded earlier, the NL3 parameter set --- predicting both a stiff
equation of state for symmetric nuclear matter ($K\!=\!271$~MeV) and
a stiff symmetry energy $R_{n}\!-\!R_{p}\!=\!0.28$~fm --- overpredicts
the centroid energy of the GMR in ${}^{90}$Zr and underpredicts the
peak position of the IVGDR in in ${}^{208}$Pb. (Note that a symmetry
energy that rises rapidly with energy predicts low values for the
symmetry energy at the densities probed by the isovector dipole
mode). In contrast, the good agreement between FSUGold and experiment
is due to the addition of the two extra parameters ($\zeta$ to reduce
the value of $K$ and $\Lambda_{\rm v}$ to soften the symmetry
energy). Note that an additional softening of the symmetry energy may
further improve the agreement with experiment. Thus, our present
prediction of $R_{n}\!-\!R_{p}\!=\!0.21$~fm could be regarded as an
upper bound.  This smaller value for the neutron skin in ${}^{208}$Pb,
generated from the softer symmetry energy, is significant as it brings
covariant meson-baryon models closer to nonrelativistic predictions
based on Skyrme parameterizations~\cite{Furnstahl:2001un}. The Parity
Radius Experiment (PREX) at the Jefferson Laboratory should provide a
unique observational constraint on the density dependence of the
symmetry energy.

  \begin{table}
  \begin{tabular}{|c|c|c|c|c|}
    \hline
    Nucleus & Observable & Experiment & NL3 & FSUGold \\
    \hline
    \hline
    ${}^{208}$Pb & GMR   (MeV) & $14.17\pm0.28$ & $14.32$ & $14.04$  \\
    ${}^{90}$Zr  & GMR   (MeV) & $17.89\pm0.20$ & $18.62$ & $17.98$  \\
    ${}^{208}$Pb & IVGDR (MeV) & $13.30\pm0.10$ & $12.70$ & $13.07$  \\
    \hline
  \end{tabular}
 \caption{Centroid energies for the breathing mode in ${}^{208}$Pb and
          ${}^{90}$Zr, and the peak energy for the IVGDR in
          ${}^{208}$Pb.  Experimental data are extracted from
          Refs.~\cite{You99_PRL82} and~\cite{Rit93_PRL70}.}
  \label{Table2}
 \end{table}

Having constructed the accurately calibrated FSUGold parameter set
with an equation of state (EoS) that is considerably softer than NL3,
we now examine some of its predictions for a few neutron-star
observables. The structure of spherically-symmetric neutron stars in
hydrostatic equilibrium is described by a solution to the
Tolman-Oppenheimer-Volkoff (TOV) equation, that may be expressed as 
a coupled set of first-order differential equations of the following 
form:
\begin{subequations}
 \begin{align}
   & \frac{dP}{dr}=-G\,\frac{{\cal E}(r)M(r)}{r^{2}}
         \left[1+\frac{P(r)}{{\cal E}(r)}\right]
         \left[1+\frac{4\pi r^{3}P(r)}{M(r)}\right]
         \left[1-\frac{2GM(r)}{r}\right]^{-1} \;, 
         \label{TOVa}\\
   & \frac{dM}{dr}=4\pi r^{2}{\cal E}(r)\;.
         \label{TOVb}
 \end{align}
 \label{TOV}
\end{subequations}
In the above equations $G$ is Newton's universal gravitational constant,
while $P(r)$, ${\cal E}(r)$, and $M(r)$ represent the pressure, energy 
density, and enclosed-mass profiles of the star, respectively. The
second of the two equations [Eq.~(\ref{TOVb})] is the differential form
for the definition of the mass enclosed $M(r)$ up to a radius $r$. That 
is,
\begin{equation}
 \frac{dM}{dr}=4\pi r^{2}{\cal E}(r)  
 \quad\iff\quad
 M(r)=\int_{0}^{r} 4\pi x^{2}{\cal E}(x) dx \;. 
\end{equation}
The dynamical information is contained in the second of the two equations 
[Eq.~(\ref{TOVa})] that embodies the notion of hydrostatic equilibrium,
namely, the gravitational attraction on any mass element of the star
must be compensated exactly by the pressure gradient
($dP/dr\!<\!0$). The last three terms on the right-hand side of this
equation (enclosed in brackets) incorporate corrections from general
relativity on to the simple Newtonian dynamics~\cite{Weinberg:1972}.
Incorporating these three correction terms is critical, as typical escape 
velocities from neutron stars are of the order of half the speed of light. 
That is,
\begin{equation}
  v_{\rm esc}/c=\sqrt{\frac{3M}{R}}\simeq 1/2 \;, 
\end{equation}
where in the above expression the mass of the star ($M$) is measured 
in solar masses and the radius ($R$) in kilometers.

\begin{figure}[ht]
\vspace{0.50in}
\includegraphics[width=4in,angle=0]{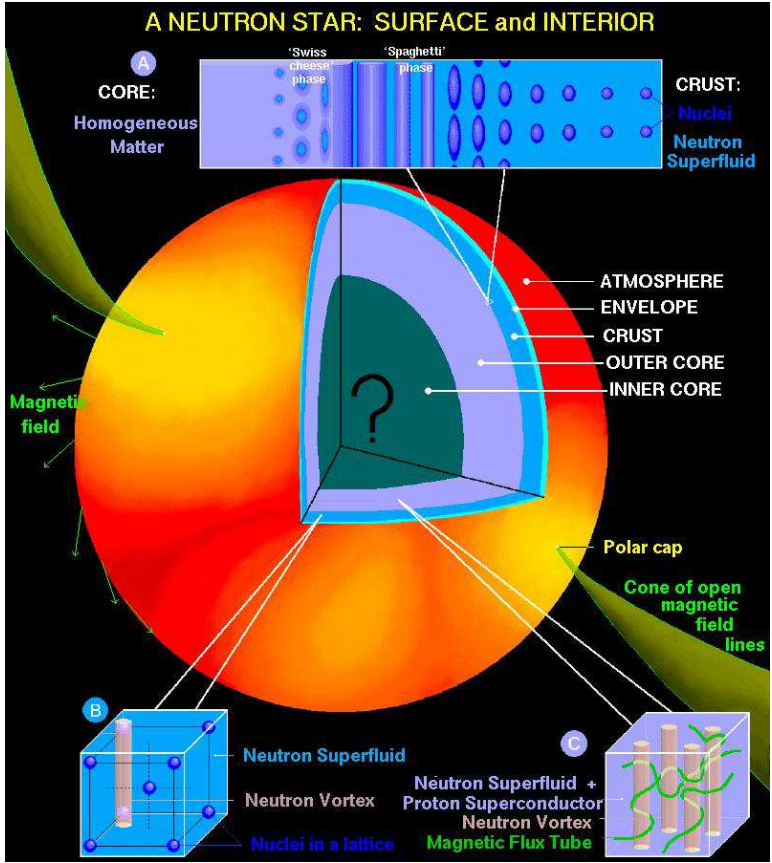}
\caption{State-of-the-art rendition of the structure of a neutron 
         star.}
\label{Fig2}
\end{figure}

The TOV equations, together with their associated boundary conditions,
{\it i.e.,} $P(r\!=\!0)\!=\!P_{c}$ and $M(r\!=\!0)\!=\!0$, are
incomplete without an equation of state $P = P({\cal E})$ to specify
the relation between the pressure and the energy density of the
system.  Thus, various properties of neutron stars, such as their
masses, radii, and temperature, probe directly the dynamics of
neutron-rich matter over a wide range of density. Thus, if general
relativity is assumed valid --- a very modest and safe assumption ---
then the structure of a neutron star is solely determined by the EoS
of neutron-rich matter in beta equilibrium.

\begin{figure}[ht]
\vspace{0.50in}
\includegraphics[width=5in,angle=0]{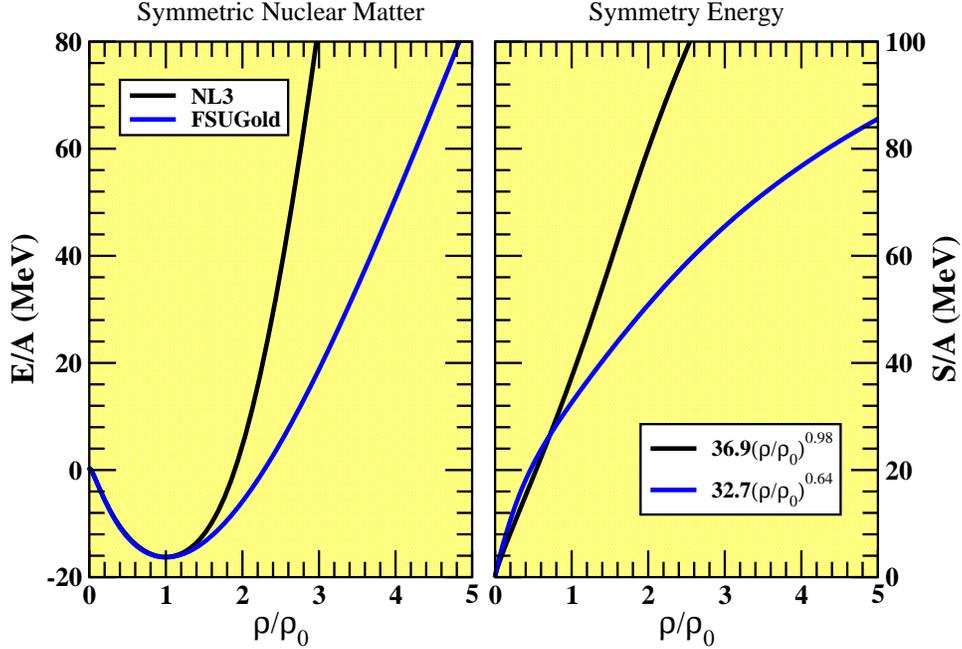}
\caption{Equation of state for symmetric nuclear matter (left panel)
         and the density dependence of the symmetry energy (right
         panel) for the NL3 and FSUGold parameter sets. The inset on
         the right-hand panel is a fit over the low-density region 
         of the predicted density dependence of the two models; 
         see text for details.}
\label{Fig3}
\end{figure}

For the uniform liquid phase in the mantle (or outer core) we assume
an EoS consisting of neutrons, protons, electrons, and muons in beta
equilibrium. Moreover, we assume that this description remains valid
in the high-density inner core. Thus, transitions to exotic phases,
such as meson condensates, hyperonic matter, and/or quark matter, are
are not considered here. Yet at the lower densities of the inner crust
the uniform system becomes unstable against density fluctuations, as
it becomes energetically favorable to separate the system into regions
of high- and low-density matter. In this nonuniform region the system
may consists of a variety of complex structures, such as spherical,
cylindrical, rods, plates, {\it etc} --- collectively known as {\it
nuclear pasta}~\cite{Ravenhall:1983uh,Hashimoto:1984}. While
microscopic calculations of the nuclear pasta are now becoming
available~\cite{Watanabe:2005qt,Horowitz:2004yf,Horowitz:2004pv,Horowitz:2005zb},
it is premature to incorporate them in our calculation. Hence,
following the procedure adopted in Ref.~\cite{Carriere:2002bx}, a
simple polytropic equation of state is used to interpolate between the
outer crust~\cite{Baym:1971pw} and the uniform liquid. For an accurate
rendition of the expected structure of a neutron star see
Fig.~\ref{Fig2}.  (This figure, courtesy of Dany Page, may be found at
{\tt http://www.astroscu.unam.mx/neutrones/NS-Picture/NS-Picture.html}).

In Fig.~\ref{Fig3} we display the equation of state for symmetric 
nuclear matter (left panel) and the symmetry energy (right panel) 
for the NL3 and FSUGold parameter sets. Note that the equation of
state for pure neutron matter is to an excellent approximation
equal to the sum of the two:
\begin{equation}
  E_{\rm PNM}(\rho) \simeq E_{SNM}(\rho) + S(\rho) \;. 
  \label{EPNM}
\end{equation}
Symmetric nuclear matter saturates for both parameter sets at a baryon
density of $\rho_{0}\!=\!0.148~{\rm fm}^{-3}$ and a binding energy per
nucleon of $E/A\!-\!M\!=\!-16.2$~MeV. Yet the addition of a quartic
vector-meson coupling $\zeta$ yields a significant softening of the
equation of state. Indeed, the nuclear matter incompressibility gets
reduced from the NL3 value of $K\!=\!271$~MeV to $K\!=\!230$~MeV for
the FSUGold set. As we shall see, differences in the predictions of
various neutron-star properties between these two models are also
significant.

Large differences in the density dependence of the symmetry energy are
also predicted by these two models. An often used parametrization of
the density dependence of the symmetry energy is given
by~\cite{Chen:2004si,Chen:2005ti,Shetty:2005qp,Ono:2005vv,Shetty:2006hh}:
\begin{equation}
  S/A=S_{0}\left(\frac{\rho}{\rho_{0}}\right)^{\gamma} \;,
 \label{SymmFit}
\end{equation}  
where $S_{0}$ is the value of the symmetry energy per nucleon at
saturation density and gamma is the parameter that quantifies its
density dependence. A fit to the density dependence of the symmetry
energy over the $0\!\le\!\rho\!\le\!\rho_{0}$ region yields the
following values for the two models:
\begin{equation}
 S_{0}=
 \begin{cases} 
   36.9~{\rm MeV} & {\rm for \, NL3,}     \\
   32.7~{\rm MeV} & {\rm for \, FSUGold,}
 \end{cases}
 \quad {\rm and} \quad
 \gamma=
 \begin{cases} 
   0.98 & {\rm for \, NL3,}     \\
   0.64 & {\rm for \, FSUGold.}
 \end{cases}
\end{equation}
As alluded earlier, the larger the value of $\gamma$ the larger the 
neutron skin of ${}^{208}$Pb. Indeed, an empirical fit to a large 
number of mean-field models yields~\cite{Horowitz:2006iv}:
\begin{equation}
  R_{n}-R_{p} \simeq (0.22\gamma + 0.06)~{\rm fm} \simeq
 \begin{cases} 
   0.28~{\rm fm} & {\rm for \, NL3,}     \\
   0.21~{\rm fm} & {\rm for \, FSUGold.}
 \end{cases}
 \label{SkinvsGamma}
\end{equation}

\begin{figure}[ht]
\vspace{0.50in}
\includegraphics[width=5in,angle=0]{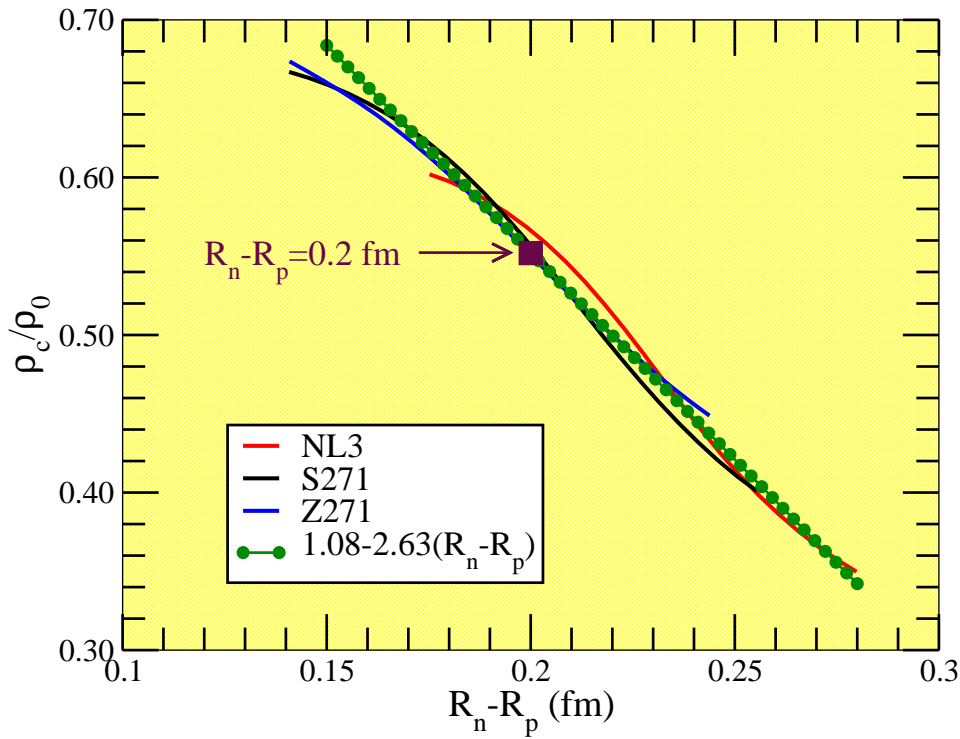}
\caption{Transition density from the uniform mantle to the
         non-uniform crust as a function of the neutron skin
         of ${}^{208}$Pb. Different models are used to show
         the largely model independent relation between these
         two quantities.}
\label{Fig4}
\end{figure}

At low densities the uniform system becomes unstable against density
fluctuations. That is, at low densities it becomes energetically
favorable for the uniform system to separate into regions of high and
low densities.  Results for the transition density from the uniform
mantle to the nonuniform crust as a function of the neutron skin in
${}^{208}$Pb are displayed in Fig~\ref{Fig4}. Various models are used
to show the nearly model-independent relation between these two
seemingly distinct observables. The figure displays an inverse
correlation between the neutron-skin and the transition density found
in Ref~\cite{Horowitz:2000xj}. This correlation suggests that models
with a stiff equation of state (such as NL3) predict a low transition
density, as it becomes energetically unfavorable to separate nuclear
matter into regions of high and low densities. Finally, this
``data-to-data'' relation illustrates how an accurate and
model-independent determination of the neutron skin of ${}^{208}$Pb at
the Jefferson Laboratory --- assumed here purely on theoretical biases
to be $R_{n}\!-\!R_{p}\!=\!0.20$~fm --- would determine an important
neutron-star observable.

A mechanism that may also benefit from an accurate determination of
the neutron radius in ${}^{208}$Pb is the cooling of a neutron star.
Proto-neutron stars are created hot in supernova explosions but then
cool rapidly through neutrino emission~\cite{Pethick:1992}.
Indeed, 99\% of the energy released in a supernovae explosion is
carried away by neutrinos. That proto-neutron stars are born with very
high temperatures was inferred from the few detected neutrinos from
SN1987A that suggest a neutrinosphere temperature as high as 5
MeV~\cite{Jegerlehner:1996kx}. Initially, the proto-neutron stars
will cool rapidly via the direct URCA process --- an efficient cooling
mechanism that consists of neutron beta decay followed by electron
capture~\cite{Lattimer:1991ib}:
\begin{subequations}
 \begin{align}
   & n \rightarrow p + e^{-} + \bar{\nu}_{e} \;, \label{DURCAa}\\
   & p + e^{-} \rightarrow n + \nu_{e}       \;. \label{DURCAb}
 \end{align}
 \label{DURCA}
\end{subequations}
As the proto-neutron star material becomes neutron rich --- a
process known as ``neutronization'' --- the direct URCA process 
ceases and is replaced by the modified URCA reaction:
\begin{equation}
  n + n \rightarrow n+ p + e^{-} + \bar{\nu}_{e} \;. 
 \label{MURCA}\\
\end{equation}
The modified URCA process, however, is relatively slow as a second 
nucleon is necessary to conserve both energy and momentum at the
Fermi surface~\cite{Page:2004fy}.  Incidentally, the term {\it URCA}
was coined by an Ukrainian --- George Gamow. After Mario Schoenberg and
George Gamow went gambling at the now defunct URCA Casino in Rio de
Janeiro, Gamow was so impressed by the roulette table where money just
disappeared that he said: ``well, the energy disappears in the nucleus
of the supernova as quickly as the money disappears at that roulette
table''. I have also been told that in Russian slang {\it URCA} can
also mean a pickpocket --- an individual that can steal your money in
a matter of seconds.

Recent X-ray observations of the neutron star in
3C58~\cite{Slane:2004jn}, Vela~\cite{Romani:2005sj}, and
Geminga~\cite{Halpern:1997} indicate low surface
temperatures. Moreover, the low quiescent luminosity in the
transiently accreting binaries KS 1731-260~\cite{Wijnands:2002fy} and
Cen X-4~\cite{Colpi:2000jc} suggest rapid cooling. As X-ray
observatories progress and our knowledge of neutron-star atmospheres
and ages improve, additional ``cold'' neutron stars may be
discovered. Such low surface temperatures appear to require enhanced
cooling from reactions that proceed faster than the modified URCA
process of Eq.~(\ref{MURCA}).

Enhanced cooling may occur via the weak decay of additional hadrons
such as pion or kaon condensates, hyperons, and/or quark
matter~\cite{Pons:2000xf,Jaikumar:2001hq,Yakovlev:2004iq}. Yet the
most conservative enhanced-cooling scenario is the direct URCA process
of Eq.~(\ref{DURCA}). This mechanism is not ``exotic'' as it only
requires protons, neutrons, electrons, and muons --- constituents
known to be present in dense matter. However, to conserve energy and
momentum at the Fermi surface the sum of the Fermi momenta of the
protons plus that of the electrons must be greater than the neutron
Fermi momentum. This requires a relatively large proton fraction
$Y_{p}\!\equiv\!Z/A$. As the symmetry energy represents a penalty
imposed on the system for departing from the symmetric $N\!=\!Z$
limit, the proton fraction is highly sensitive to the density
dependence of the symmetry energy.  In essence, the stiffer the
symmetry energy the larger the proton fraction --- as it is
energetically unfavorable for the proton fraction to depart
significantly from its symmetric $Y_{p}\!=\!1/2$ value.

\begin{figure}[ht]
\vspace{0.50in}
\includegraphics[width=5in,angle=0]{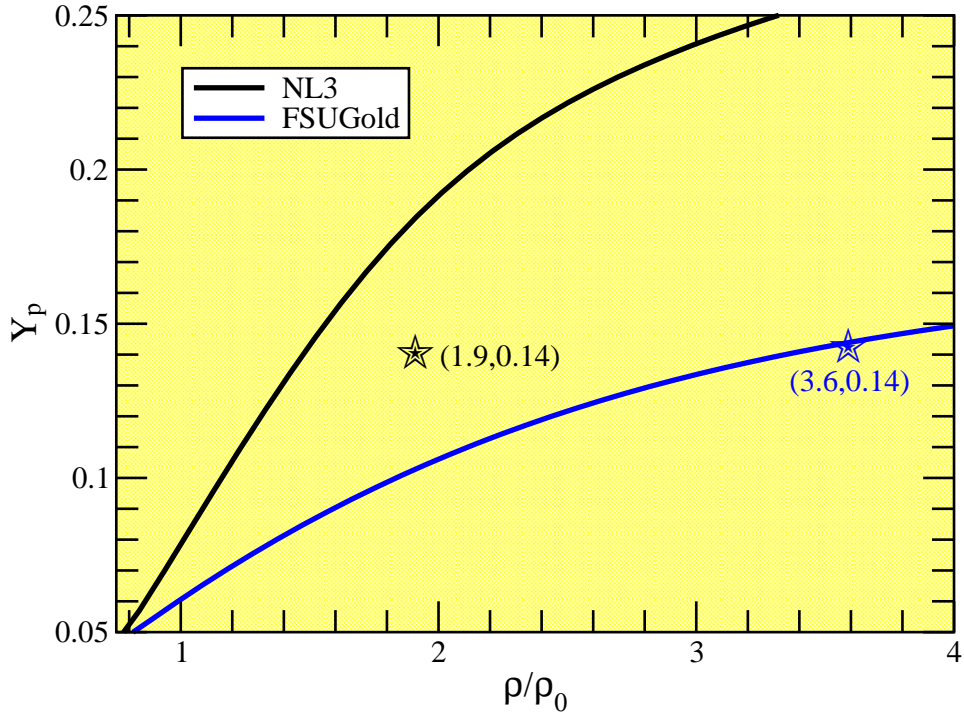}
\caption{Proton fraction as a function of baryon density for the 
         NL3 and FSUGold parameter sets. The ``stars'' indicate
         the central density and the minimum value of the proton
         fraction necessary for the direct URCA process to occur
	 in a 1.4 solar-mass neutron star.}
\label{Fig5}
\end{figure}

In Fig.~\ref{Fig5} we display the proton fraction $Y_{p}$ as a function
of baryon density for the NL3 and FSUGold parameter sets. The location
of the symbols (``stars'') and the quantities enclosed in parenthesis
indicate the central density and the minimum value of the proton
fraction necessary for the direct URCA process to be effective in a
$M\!=\!1.4 M_{\odot}$ neutron star. In particular, if the solid curve
passes above the symbol, then the direct URCA process is allowed in
such a neutron star.  Several features of this plot are worth focusing
on. First, because of its considerable stiffer symmetry energy, the
proton fraction predicted by the NL3 set increases much more rapidly
than for the FSUGold model. Second, not only does the symmetry energy
increases more rapidly in the case of the NL3 set, but so does the
equation of state for symmetric nuclear matter (see
Fig.~\ref{Fig3}). As a result, the pressure required to support a
$M\!=\!1.4 M_{\odot}$ neutron star against gravitational collapse is
reached at a much higher density in the FSUGold model than in the NL3
set: $\rho_{c}\!=\!3.6\,\rho_{0}$ for FSUGold {\it vs}
$\rho_{c}\!=\!1.9\,\rho_{0}$ for NL3. It is precisely this fact --- the
much higher density reached in the case of the FSUGold parameter set
--- that allows a $M\!=\!1.4 M_{\odot}$ to be cooled by the direct
URCA process in spite of its soft symmetry energy.  Yet, either a
softer equation of state for symmetric nuclear matter or a softer
symmetry energy may preclude the enhanced cooling of a $M\!=\!1.4
M_{\odot}$ neutron star by the (nucleon) direct URCA process. If such
is the case, then the enhanced cooling of a $M\!=\!1.4 M_{\odot}$
neutron star may provide strong evidence in favor of exotic matter
({\it e.g.,} ``quark matter'') in the core of a neutron star.  We
close this section by listing in Table~\ref{Table3} the predictions of
both models for a variety of neutron-star observables.

  \begin{table}
  \begin{tabular}{|c|c|c|}
    \hline
     Neutron-Star Observable & NL3 & FSUGold \\
    \hline
    \hline
     $\rho_{c}$~(fm$^{-3}$)          & $0.052$ & $0.076$  \\
     $R$~(km)                        & $15.05$ & $12.66$  \\
     $M_{\rm max}(M_{\odot})$        & $ 2.78$ & $ 1.72$  \\
     $\rho_{_{\rm URCA}}$~(fm$^{-3}$)& $ 0.21$ & $ 0.47$  \\
     $M_{\rm URCA}(M_{\odot})$       & $ 0.84$ & $ 1.30$  \\
     $\Delta M_{\rm URCA}$           & $ 0.38$ & $ 0.06$  \\
    \hline
  \end{tabular}
  \caption{Predictions for a few neutron-star observables. The various
           quantities are as follows: $\rho_{c}$ is the transition
           density from nonuniform to uniform neutron-rich matter
           matter, $R$ is the radius of a 1.4 solar-mass neutron star,
           $M_{\rm max}$ is the limiting mass, $\rho_{_{\rm URCA}}$ is
           the threshold density for the direct URCA process, $M_{\rm URCA}$
           is the minimum mass neutron star that may cool down by the
           direct URCA process, and $\Delta M_{\rm URCA}$ is the mass
           fraction of a 1.4 solar-mass neutron star that supports
           enhanced cooling by the direct URCA process.}
  \label{Table3}
 \end{table}

\section{Conclusions}
\label{Sec:Conclusions}

In conclusion, a new accurately calibrated relativistic model
(``FSUGold'') has been fitted to the binding energies and charge radii
of a variety of magic nuclei. In this regard, the new parametrization
is as successful as the NL3 set which has been used here as a useful
paradigm. In particular, symmetric nuclear matter saturates at a Fermi
momentum of $k_{\rm F}\!=\!1.30~{\rm fm}^{-1}$ (corresponding to a
baryon density of $0.15~{\rm fm}^{-3}$) with a binding energy per
nucleon of $B/A\!=\!-16.30$~MeV.  Further, by constraining the FSUGold
parameter set by a few nuclear collective modes, we obtain a
nuclear-matter incompressibility of $K\!=\!230$~MeV and a neutron skin
thickness in ${}^{208}$Pb of $R_{n}-R_{p}\!=\!0.21$~fm. While the
description of the various collective modes imposes additional
constraints on the EoS at densities around saturation density, the
high-density component of the EoS remains largely unconstrained. Thus,
we made no attempts at constraining the EoS at the supranuclear
densities of relevance to neutron-star physics.  Rather, we simply
explored the consequences of the new parametrization on a variety of
neutron star observables and eagerly await high-quality data that will
constrain the high-density component of the EoS. In particular, we
found a limiting neutron-star mass of $M_{\rm max}\!=\!1.72
M_{\odot}$, a radius of $R\!=\!12.66$~km for a $M\!=\!1.4 M_{\odot}$
neutron star, and no direct URCA cooling in neutron stars with masses
below $M\!=\!1.3 M_{\odot}$.  It is interesting to note that recent
observations of pulsar-white dwarf binaries at the Arecibo observatory
suggest a pulsar mass for PSRJ0751+1807 of $M\!=\!2.1^{+0.4}_{-0.5}
M_{\odot}$ at a 95\% confidence level~\cite{Nice:2005fi}.  If this
observation could be refined, not only would it redefine the
high-density behavior of this (and many other) EoS, but it could
provide us with a precious boost in our quest for the equation of
state.

\begin{acknowledgments}
The author is extremely grateful to the organizers of the
NPAE-Kyiv2006 conference for their warmth and hospitality.
This work was supported in part by DOE grant DE-FG05-92ER40750.
\end{acknowledgments}

\vfill\eject

\end{document}